# Millimeter-sized magnetic domains in perpendicularly magnetized ferrimagnetic Mn$_4$N thin films grown on SrTiO$_3$

Toshiki Gushi[1,2], Laurent Vila[2], Olivier Fruchart[2], Alain Marty[2], Stefania Pizzini[3], Jan Vogel[3], Fumiya Takata[1], Akihito Anzai[1], Kaoru Toko[1], Takashi Suemasu[1,a)], and Jean-Philippe Attané[2,b)]

[1] *Institute of Applied Physics, Graduate School of Pure and Applied Sciences, University of Tsukuba, Tsukuba, Ibaraki 305-8573, Japan*

[2] *Univ. Grenoble Alpes, CEA, CNRS, INAC-Spintec, 38000 Grenoble, France*

[3] *Univ. Grenoble Alpes, CNRS, Institut NEEL, 38000 Grenoble, France*

a) Electronic mail: suemasu@bk.tsukuba.ac.jp

b) Electronic mail: jean-philippe.attane@cea.fr

## Abstract

The use of epitaxial layers for domain wall-based spintronic applications is often hampered by the presence of pinning sites. Here, we show that when depositing Mn$_4$N(10 nm) epitaxial films, the replacement of MgO(001) by SrTiO$_3$(001) substrates allows minimizing the misfit, and to obtain an improved crystalline quality, a sharper switching, a full remanence, a high anisotropy and remarkable millimeter-sized magnetic domains, with straight and smooth domain walls. In a context of rising interest for current-induced domain wall motion in rare-earth ferrimagnets, we show that Mn$_4$N/SrTiO$_3$, which is rare-earth free, constitutes a very promising ferrimagnetic system for current-induced domain wall motion.

**Text**

Current-induced domain wall motion (CIDWM) is actively studied theoretically[1-4] and experimentally[5-7], because it raises new questions concerning the interplay between spin and charge currents, and in order to prospect for new spintronic devices such as non-volatile memories based on domain wall (DW) motion[8]. During one decade, the spin-transfer torque resulting from the current flow within the magnetic material was the only physical effect known to allow domain wall motion. The discovery of spin-orbit torques (SOT) in multilayer systems[9,10] brought more versatility and a greater efficiency to CIDWM, and largely shifted the focus of the community towards spinorbitronics[11].

Simulations and experiment[12-14] show that in perpendicularly magnetized ferrimagnets the critical current densities required to induce domain-wall motion by spin transfer torque or by SOT is greatly reduced, because of the low value of the magnetization. Minimizing the operating currents allows obtaining a lower energy consumption and areal footprint[15]. Thus, the combination of both STT/SOT and ferrimagnets is a very attracting pathway for creating spintronic devices based on DW motion or on nanomagnet switching. For example, Je *et al*. reported a large enhancement of the SOT efficiency in a CoTb/Pt stack with a compensated alloy composition[16] ($3.7 \times 10^{-13}$ Tm$^2$/A, three times larger than that of Ta/CoTb[17], and five times larger than that of Pt/Co/AlO$_x$[18]).

Until now, most perpendicular ferrimagnetic materials used for DW motion experiments have been rare-earth based, with potential concerns about material criticality in the context of the rising demand for these elements. Anti-perovskite-type nitrides with 3*d* transition metals are rare-earth free, and could thus be interesting candidates for the replacement of the rare-earth based systems. Among them, Mn$_4$N has a high Curie temperature $T_C$ of 745 K[19], a small spontaneous magnetization $M_S$ (100 kA/m) and a strong perpendicular magnetic anisotropy (PMA), with $\mu_0H_K$ values over 2.5 T[20,21]. Mn$_4$N samples with PMA have been successfully grown on several kinds of substrates, such as glass[22], Si(100)[23,24], SiC[25], MgO[20,21,26,27] and SrTiO$_3$(STO) [20,21].

In the following, we study 10-nm-thick Mn$_4$N epitaxial thin films deposited on MgO(001) and STO(001) substrates by molecular beam epitaxy (MBE). Using structural characterizations, magneto-transport measurements, vibrating sample magnetometer (VSM) and magnetic imaging techniques, we study the influence of the substrate on the magnetic properties, showing that the use of STO substrates allows obtaining astonishing DW properties, with seemingly very few pinning sites, and a domain size in the millimeter range.

SiO$_2$(3 nm)/Mn$_4$N(10 nm) layers have been deposited on MgO(001) and STO(001) substrates at 450 °C, using a MBE system with an ion-pump (10$^{-7}$ Pa), equipped with a high-temperature Knudsen cell for Mn and a radio-frequency (RF) N$_2$ plasma[20,21]. To prevent oxidation, the samples have been capped with 3-nm-thick SiO$_2$ layers.

The crystalline quality of the Mn$_4$N layer has been characterized by 20-kV reflection high-energy electron diffraction (RHEED), and x-ray diffraction (XRD) with Cu K$_\alpha$ radiation. Figure 1 presents the out-of-plane ($\omega - 2\theta$) and in-plane ($\phi - 2\theta_\chi$) XRD with the RHEED patterns of the Mn$_4$N layers grown on MgO and STO substrates, respectively. $\omega$-scan rocking curves are shown in the same figure for (e) Mn$_4$N 002 on MgO, (f) Mn$_4$N 004 on STO. For Mn$_4$N/MgO, the XRD peaks of the film and substrate are well split, which allows extracting information on the magnetic layer. On the contrary, the diffraction peak of Mn$_4$N 002 is so close to STO 002 that we need to analyze the higher-order Mn$_4$N 004 rocking curves to separate each peak. These rocking curves have been fitted by a Lorentzian function.

A common feature of both systems is the observation of streaky RHEED patterns and of XRD peaks of *c*-axis-oriented $Mn_4N$ such as 001 or 002. Both these observations are proofs of the epitaxial growth of the $Mn_4N$ film. However, the width of the rocking curves are obviously different, indicating that $Mn_4N$ films are much better textured when deposited on STO rather than on MgO. This result is similar to what has been observed for $Fe_4N$ films[28]. The superlattice reflections in the RHEED pattern, together with the 001 peak in the XRD pattern, indicate the good long-range ordering and the presence of the N atom at the body center of the fcc-Mn lattice. X-ray reflectometry provides also an accurate measurement of the magnetic films thicknesses, giving 8.8 nm for the $Mn_4N$/MgO sample and 9.4 nm for the $Mn_4N$/STO sample.

Magneto-transport properties, such as the magnon-magnetoresistance (MMR)[29,30] and the Anomalous Hall Effect (AHE), have been measured by the Van der Pauw method for $Mn_4N$ blanket layers at room temperature, using lock-in techniques at 210 Hz. The resistivity at room temperature is 187 μΩ.cm for the $Mn_4N$/MgO sample, and 181 μΩ.cm for the $Mn_4N$/STO sample. As shown in Figs. 2(a) and (b), the magnetoresistance is dominated by the MMR, with a behavior typical of samples with a strong uniaxial anisotropy.

Figures 2 (c) and (d) show the hysteresis loops measured by AHE. The AHE angle $\rho_{xy}/\rho_{xx}$ is high (−2%), in line with previous reports on $Mn_4N$[31,32]. The coercive decreases slightly in the $Mn_4N$/MgO sample, knowing that these high coercive fields are due to the small saturation magnetization. While the $Mn_4N$/MgO sample shows a smooth hysteresis loop, the magnetization of the $Mn_4N$/STO sample switches very sharply, with a full remanence at zero field. Despite these differences, it is striking that the spontaneous magnetization and all the transport quantities of the two systems are very similar, which indicating that the materials are intrinsically alike. The structural characterizations suggest however that the structural disorders due to the crystalline defects are different, and thus that the increased disorder is responsible for the higher coercivity and slanted loop of Mn$_4$N/MgO samples.

Figures 2(e) and 2(f) present the out-of-plane and in-plane magnetization curves, obtained by VSM-SQUID up to 4 and 6 T, respectively, of Mn$_4$N/MgO and Mn$_4$N/STO. From the out-of-plane hysteresis curves, we found the $M_S$ of Mn$_4$N/MgO and Mn$_4$N/STO to be 118 and 105 kA/m, respectively. From the in-plane loop of Mn$_4$N/STO, the anisotropy field $H_K$ can be estimated to be 4 T. The uniaxial anisotropy $K_u$ was calculated to be $1.1 \times 10^5$ J/m$^3$, from the integration of the area enclosed between the in-plane and out-of-plane magnetization curves, and taking into account the demagnetization energy.

Let us discuss now the magnetic domain configuration in Mn$_4$N thin films. The theoretical equilibrium domain size in Mn$_4$N layers results from the balance between the dipolar energy and the DW energy. With $M_S$=105 kA/m, $K_u$ =110 kJ/m$^3$, and an exchange stiffness of $A$=15 pJ/m (using a rough estimation from the Curie temperature [33], the DW width can be calculated to be $\pi\sqrt{A/K_u} = 37$ nm. The resulting equilibrium domain size for a 10-nm-thick Mn$_4$N film, calculated using the analytical model of ref. [34], is of several km. This indicates that the demagnetizing field is negligible because of the small $M_S$, and that in practice the domain size and shape shall be rather determined by DW pinning on extrinsic defects[35].

Let us examine the domain pattern in Mn$_4$N layers on MgO and STO substrates. For each kind of sample, the magnetic domain configuration has been observed by magnetic force microscopy (MFM) and/or magneto-optical Kerr effect (MOKE) microscopy, depending on the typical length scale found for the magnetic domains. The mean domain period $d_p$ has been estimated using a 2D fast Fourier transformation method [36]. The observation has been performed both for the as-deposited state, and after partial magnetization reversal at the coercive field. Figures 3 (a) and (b) show the magnetic domain configuration before any application of magnetic field, in a state which is often considered to be close to the equilibrium state. In Mn$_4$N/MgO samples, magnetic domains are small ($d_p$ = 0.28 µm), far below the theoretical equilibrium width. The irregular shape of the DWs is characteristic of a strong disorder[35]. On the contrary, the magnetic domain size in the Mn$_4$N/STO sample is two orders of magnitude larger: although it remains smaller than the theoretical equilibrium domain width, $d_p$ is as large as 20 µm, with smooth magnetic DWs. This indicates that there is a much lower influence of the extrinsic disorder in the Mn$_4$N/STO sample. The domain width value is comparable with that of ultrathin CoFeB(1.1 nm)/MgO(1) and Pt(2.4)/Co(0.27) system, with $d_p$ values of respectively 14 and 6.5 µm obtained after thermal demagnetization[36,37].

Figure 3(c) shows the Hall signal of Mn$_4$N/STO during a typical partial reversal process. Figure 3 (d) and (e) present the domain configurations of both samples in partially reversed states. In the case of Mn$_4$N/MgO, the structure of the domains is almost the same as in the as-deposited state, while the domain structure in Mn$_4$N/STO evolved from a micron-sized configuration to a millimeter-sized one, a value never reported, to our knowledge, in any perpendicularly magnetized thin films. MOKE images show the presence of very few nucleation centers, showing that the magnetization switching occurs by a nucleation followed by an easy propagation. Note that this observation is consistent with the fact that the hysteresis loop is square.

The differences between the two systems are striking, concerning both their hysteresis loops, and their domain shapes and sizes. The XRD data and the magnetic properties suggest that these differences arise from the crystalline quality of the samples, which is higher for the Mn$_4$N/STO samples. The physical origin could arise from the difference in the strain relaxation process: there is a large lattice misfit $f = (a_{film} - a_{sub})/a_{sub}$ at the Mn$_4$N/MgO interface (−7.6%), whereas the misfit at the Mn$_4$N/STO interface is only −0.4%. While the existence of misfit dislocations at Mn$_4$N/MgO interface has already been observed by transmission electron microscopy[26], other strain-relaxation through-dislocations of micro-grain boundaries are also expected to play a major role in the magnetic behavior. This underlines that the selection of a well-matching substrate is crucial to improve the magnetic properties of Mn$_4$N layers, and notably its suitability for current-induced DW propagation.

In summary, we showed that the magnetic properties of perpendicularly magnetized Mn$_4$N layers are dramatically improved when replacing MgO substrates by STO substrates. This Mn$_4$N/STO system exhibits astonishing properties: a giant magnetic domain structure, at the millimeter length scale, with full remanence, scarce nucleation and a sharp magnetization switching. These properties, associated to a very small $M_S$ and a large PMA, underline the potential of Mn$_4$N/STO layers as a rare-earth-free ferrimagnet for CIDWM.

## ACKNOWLEDGMENTS

This work was supported in part by JSPS Grants-in-Aid for Scientific Research (A) (No. 26249037) and JSPS Fellows (Nos. 16J02879). Magnetization measurements have been performed with the help of Professor H. Yanagihara of the University of Tsukuba, and MFM images thanks to the help of Simon Le Denmat, from Institut Néel. Support from the ANR OISO and Laboratoire d'excellence LANEF in Grenoble (ANR-10-LABX-51-01) are also acknowledged.

**Figure Captions**

Fig. 1. XRD profiles of Mn$_4$N films on (a), (c), (e) MgO and (b), (d), (f) STO substrate. (a), (b) Out-of-plane XRD patterns. The insets show the stack structures. The blue marks indicate the peaks attributed to the (100)-oriented-Mn$_4$N. (c), (d) In-plane XRD patterns. The incidence angle is $\omega$=0.4°, and the scattering vector is along [100]. The insets show the RHEED images of the Mn4N layers taken along the [100] direction of each substrate. (e), (f) $\omega$-scan rocking curves for (e) Mn$_4$N 002 on MgO, and (f) Mn$_4$N 004 on STO. The black and red lines represent the raw data and the fit using a Lorentzian curve, respectively.

Fig. 2. (a), (b) Dependence of the longitudinal resistance with the perpendicularly applied magnetic field. (c), (d) Anomalous Hall effect hysteresis loops. (e), (f) Out-of-plane hysteresis loops measured by VSM. (a), (c), (e) correspond to the Mn$_4$N/MgO sample, and (b), (d), (f) to the Mn$_4$N/STO sample.

Fig. 3. (a) MFM image of the Mn$_4$N/MgO and (b) MOKE image of the Mn$_4$N/STO samples in the as-deposited state. (c) Magnetization curve illustrating the partial reversal process, monitored by anomalous Hall effect in the Mn$_4$N/STO sample. The red dot corresponds to the final magnetization state, used for (e). (d) MFM image of the Mn$_4$N/MgO sample and

(e) MOKE image of the Mn$_4$N/STO sample after partial reversal.

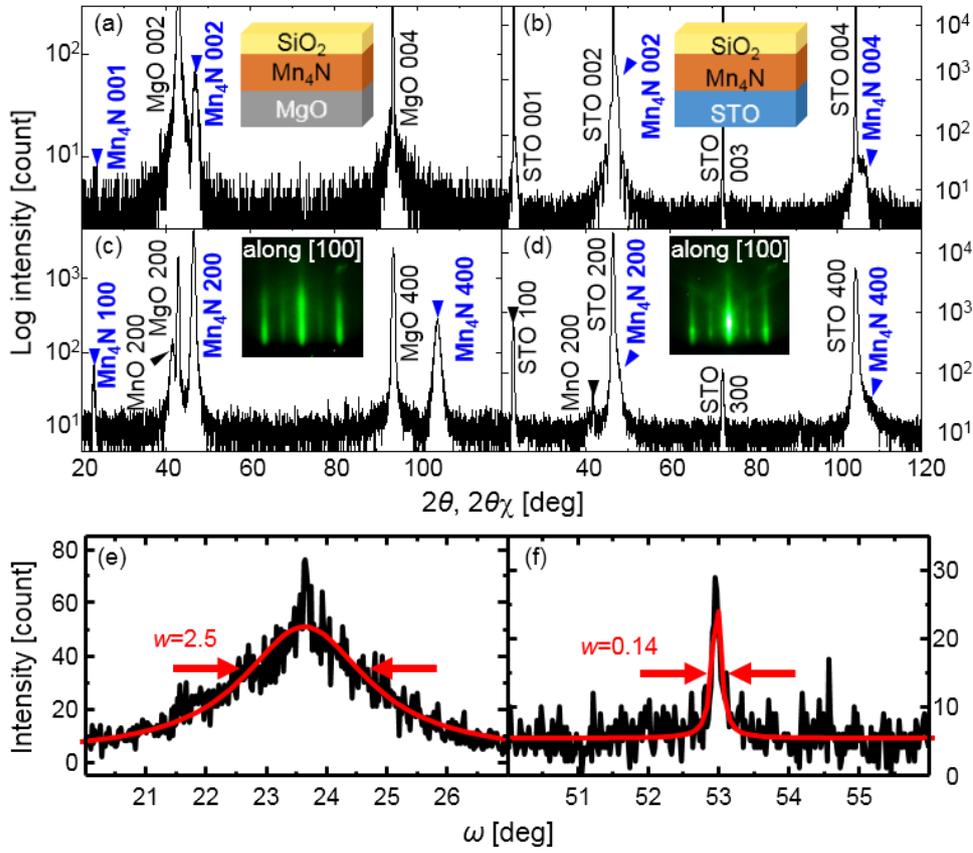

Fig.1.

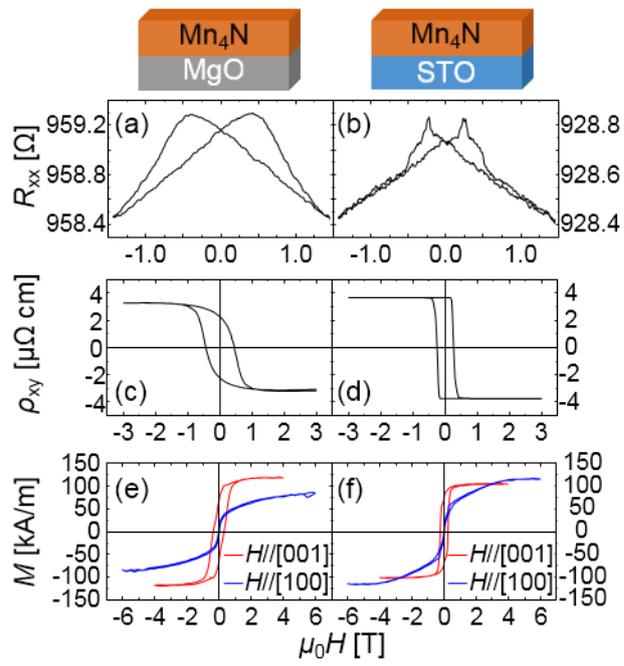

Fig.2.

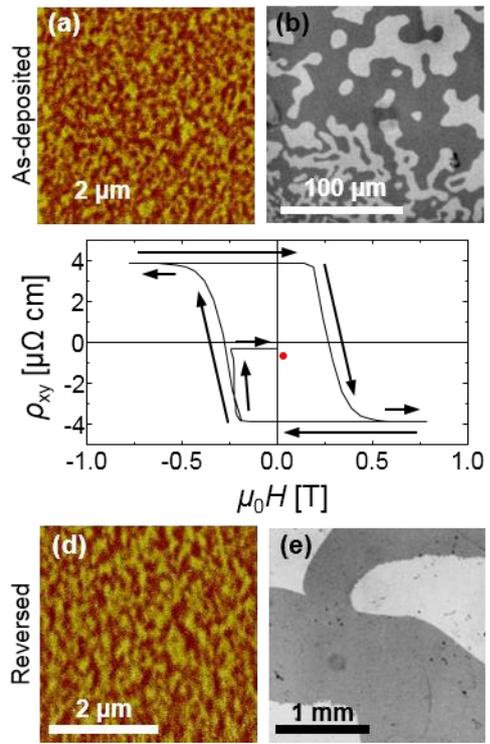

Fig.3.